\documentclass[%
aip,
amsmath,amssymb,
reprint,%
]{revtex4-1}

\usepackage{graphicx}
\usepackage{amsmath}
\usepackage{amssymb}
\usepackage{float}
\usepackage{color}
\usepackage{makecell}
\usepackage{amsfonts}
\usepackage{tabularx}
\usepackage{adjustbox}
\usepackage[figuresright]{rotating}
\restylefloat{table}
\usepackage[utf8]{inputenc}
\usepackage[T1]{fontenc}
\usepackage{mathptmx}
\usepackage{etoolbox}
\begin{document}
	
	\title {Widom insertion method in simulations with Ewald summation}
	
	\author{Amin Bakhshandeh }
	\email{bakhshandeh.amin@gmail.com}
	\affiliation{Instituto de F\'isica, Universidade Federal do Rio Grande do Sul, Caixa Postal 15051, CEP 91501-970, Porto Alegre, RS, Brazil}

	\author{Yan Levin}
	\email{levin@if.ufrgs.br}
	\affiliation{Instituto de F\'isica, Universidade Federal do Rio Grande do Sul, Caixa Postal 15051, CEP 91501-970, Porto Alegre, RS, Brazil}
	
	\begin{abstract}
		We discuss the application of Widom insertion method for calculation of the chemical potential of individual ions in computer simulations with Ewald summation.  Two approaches are considered.  In the first approach an individual ion is inserted into a periodically replicated overall charge neutral system representing an electrolyte solution.  In the second approach an inserted ion is also periodically replicated, leading to the violation of the overall charge neutrality.  This requires an introduction of an additional neutralizing background.  We find that the second approach leads to a much better agreement with the results of  grand canonical Monte Carlo simulation for the total chemical potential of a neutral ionic cluster.  
		
	\end{abstract}

	\maketitle 
	
	
	\section{Introduction}
	Ion chemical potential is an important thermodynamic quantity  which is relevant for phase equilibrium and reaction chemistry.   However measuring the chemical potential, or equivalently the solvation free energy of individual ions, is very difficult experimentally,
	requiring some specific assumptions~\cite{Latimer}. 
	On the other hand,  chemical potential of ions can be calculated approximately using theoretical methods such as Mean Spherical Approximation (MSA) or Hypernetted Chain (HNC) equation~\cite{waisman1972mean,triolo1976simple,fushiki1988hypernetted}.  Such approaches, however, are not exact and rely on specific closure relations of Ornstein-Zernike equation.  Therefore, it is desirable to have an ``exact'' method to obtain chemical potential using Monte Carlo (MC) simulations.  For systems with short range interactions there are two usual approaches:  (1)  grand canonical MC simulation (GCMC) and (2) Widom insertion method.    
	
	The simulations of Coulomb systems are significantly more complicated than those of systems with short range forces.   The long-range nature of the Coulomb potential precludes the use of simple periodic boundary conditions, requiring a periodic replication of the whole system.  Each ion, then, interacts with all the other ions inside the simulation cell and also with all the periodic replicas of all these ions.  To efficiently account for the periodicity of the replicated systems, the usual approach is to use Ewald summation methods~\cite{in2007application,delville2000electrostatic,deserno1998mesh,wang2001estimate,de2002electrostatics,bkh2019,bkh2020,ef2018,linse2005simulation,cuetos2006layering,kolafa1992cutoff,levin}.   In  thermodynamic limit the system must be charge neutral, the GCMC must, therefore, be implemented in such a way as to respect this requirement.  The simples way to do this is to insert charge neutral clusters into the simulation box.  Such approach, however,   precludes us from determining individual chemical potential of ions, allowing only calculation of the total chemical potential of a neutral cluster.  For example, in the case of $\alpha$:1 electrolyte, where $\alpha$ refers to cation valence,  we can only determine the combination $\mu_t=\mu_++\alpha \mu_-$, where $\mu_+$ and $\mu_-$ are the cation and anion chemical potentials, respectively.  Therefore, such implementation of GCMC does not provide us with an access to individual chemical potentials  $\mu_+$ and $\mu_-$, but only to $\mu_t$. We should note, however, that there is a different implementation of GCMC in which individual ions, together with their respective neutralizing background, are inserted into the simulation box~\cite{Barr}.  The difficulty in such approach is that the chemical potential of cations and anions must be carefully adjusted, so that neutrality of the simulation box is due only to ions and not  because  of an artificial background.

	
	An alternative approach which allows us to obtain individual chemical potentials of ions is the Widom insertion method~\cite{widom}.  Widom  showed that the chemical potential of a particle is related to the acceptance probability of inserting particle $N+1$  into the  system that already contains $N$ particles~\cite{adams1974chemical,shing1982chemical,frenkel1992novel,widom,groot2000mesoscopic,dullens2005widom,mladek2011pair,nezbeda1991new,boda2011analyzing,widom1978structure}:
	\begin{equation}\label{e1} 
	\mu_{ex}= -k_{B}T \ln \left< \frac{1}{V} \int ds_{N+1}  \exp(-\beta\Delta U)\right>_{N},
	\end{equation}
	where $\Delta U\equiv U(s^{N+1}) - U(s^{N})$ is the energy difference  for systems with $N$ and $N+1$ particles. The integral is easily calculate inside a canonical MC simulation by sampling the insertion probability $\exp(-\beta\Delta U)$ after the simulation with $N$ particles has fully equilibrated~\cite{svensson1988widom,maciel2018chemical,sloth1990monte,xu2021calculation,frenkel1996}.
	
	Widom's method  has been widely used for evaluating the excess chemical potential for different systems such as supercritical fluid-solid equilibria\cite{albo2003calculation,pai2014solubility}, mixture of Argon and 1-magne-4-polybutadiene~\cite{gestoso2008barrier}, binary phases~\cite{carrero2008simulation} etc.
	Widom insertion  method was also used to calculate ionic solvation free energy in atomistic simulations~\cite{hunenberger1999ewald,levy1998computer,maciel2018chemical,Panagiotopoulos}. 
	
	
	
	To use Eq.~\ref{e1} requires calculation of  $\Delta U$, which is the change in energy of the system due to addition of a test ion.  Within Ewald summation formalism there is, however, an ambiguity in the definition of  $\Delta U$.   One way to interpret $\Delta U$ as the energy due to the interaction of an extra ion with all the other ions inside the simulation cell, as well as with all the replicas of these ions.  There is no problem with violation of charge neutrality in this case since only one extra ion is added to a charge neutral system, and this ion is not replicated.  An alternative is to treat the added ion on the same footing as the other ions inside the system.   In this case both the new ion and its periodic replicas must be used  to  calculate  $\Delta U$. This will lead to the interaction of ion with its own replicas, resulting in a non-neutral macroscopic system with  diverging electrostatic energy.  To overcome this difficulty we can add a uniform neutralizing background  which is introduced simultaneously  with the inserted ion.  The background charge will be replicated together with the ion, preserving the overall charge neutrality.   This will result in an overall charge neutral system 
with extensive energy.  A priori it is not clear which one of this procedures will lead to a better approximation to the exact value of the ionic chemical potential.  We should note, however, that within minimum image approximation inclusion of neutralizing background has been found to lead to much faster convergence to the thermodynamic limit~\cite{sloth1990monte}.  In this paper we will test both Ewald summation approaches by calculating the chemical potential of cations and anions separately and then compare the resulting value of $\mu_t$ obtained using each approach with the value of $\mu_t$ calculated using GCMC.  The GCMC will provide us with a benchmark to measure the accuracy of the two Widom insertion methods for periodically replicated systems.
	
	The rest of the paper is organized as follows: In section II we briefly review the grand canonical simulation method for $\alpha$:1 electrolyte, in section III we will derive the expressions for  $\Delta U$ used in the two Widom insertion methods.  In sections IV  we will present the results of the simulations obtained using the two $\Delta U$ and compare the results with the $\mu_t$ calculated using the GCMC simulations.   Finally, in section V we will discuss the conclusions of the present work.

	\section{Grand Canonical Monte Carlo Simulation} \label{s1}
	To calculate  $\mu_t$ we can perform GCMC simulations for $\alpha$:1 electrolyte. To this end, we use a cubic simulation cell with side length $L=100~$\AA. To account for the long range Coulomb interaction we use the Ewald summation method for neutral systems~\cite{ewald1921ewald,darden1993particle} with the number of k-vectors around $600$. The system is found to reach equilibrium after  $2\times10^6$ MC steps. $20000$ samples are then used for the statistical analysis. 
	In each MC move there are three possibilities:  simple movement of ions or addition or removal of one cation and $\alpha$ anions, so as to preserve the overall charge neutrality of the system.  The  transition probability for addition of ions (from state $i$ to $j$)~\cite{frenkel1996,bakh2015,valleau1980primitive,allen2017computer}:   
	\begin{equation}
	\frac{\rho_j}{\rho_i} = \frac{V^{\alpha+1}\mathrm{e}^{-\beta U_j+\beta U_i + \beta\mu_t }}{(N_+ + 1)(N_-+\alpha)(N_-+\alpha-1)...(N_-+1)\Lambda_+^{3}\Lambda_-^{3\alpha}} \ ,
	\end{equation}
	where $V$ is the volume of the simulation cell, $N_\pm$ are the number of cations and anions, $U_i$ is the electrostatic energy of the state $i$, $\mu_t=\mu_++\alpha \mu_-$ the total chemical potential of a minimum neutral cluster, and $\Lambda_\pm$ are the thermal de Broglie wavelengths of cations and anions. The removal probability is:
	\begin{equation}
	\frac{\rho_j}{\rho_i} = \frac{\mathrm{e}^{-\beta U_j+\beta U_i-\beta\mu_t }N_+ N_-(N_- -1)...(N_--\alpha+1)\Lambda_+^{3}\Lambda_-^{3\alpha}}{V^{\alpha+1}} \ .
	\end{equation}
	We start with an empty simulation cell and specify $\mu_t$ of the reservoir.  The simulation is then run until the equilibrium is established and the average number of cations inside the simulation cell is calculate $\langle N_+\rangle$.  From this we calculate the average concentration of electrolyte $\langle c \rangle$ corresponding to a fixed value of fugacity $\exp(\beta \mu_t)/\Lambda_+^{3}\Lambda_-^{3\alpha}$. The excess part of the total chemical potential can then be calculated as $\mu_t^{ex}=\mu_t- \ln[\langle c \rangle ^{\alpha+1} \Lambda_+^{3}\Lambda_-^{3\alpha}]-\alpha \ln \alpha $.

	\section{Widom Insertion Method}  \label{s2}
	
	The difficulty with applying Widom insertion method to systems with Coulomb interactions is due to the necessity of periodic replication of the simulation box.  The electrostatic potential inside the simulation cell satisfies the Poisson equation 
	\begin{equation}\label{main}
	\nabla^2 \phi({\bf r})=-\frac{4\pi q_i}{\epsilon_w}\sum_{j=1}^{N}\sum_{n_x,n_y, n_z=-\infty}^{\infty}\delta({\pmb r}-{\pmb r}^j+n_x L \hat{\pmb x}+n_y L\hat{\pmb y} +n_z L\hat{\pmb z})
	\ ,
	\end{equation}
	where $\epsilon_w $ is the dielectric constant of water and $n$'s are integers corresponding to periodic replicas. Using the usual procedure the equation can be integrated by separating the Coulomb potential into long and short range contributions.  The long range contribution can be efficiently summed in the Fourier space, while the short range in the real space.  
	The electrostatic potential can them be written as
	\begin{eqnarray}\label{e4}
	\phi({\pmb r})&=&\sum_{{\pmb k}={\pmb 0}}^{\infty}\sum_{j=1}^{N}\frac{4\pi \text{q}^j}{\epsilon_w V |{\pmb k}|^2}\exp{[-\frac{|{\pmb k}|^2}{4\kappa_e^2}+i{\pmb k}\cdot({\pmb r}-{\pmb r}^j)]} + \nonumber \\
	&&\sum_{j=1}^{N}\sum_{{\pmb n}}\text{q}^j\frac{\text{erfc}(\kappa_e|{\pmb r}-{\pmb r}^j-L{\pmb n}|)}{\epsilon_w |{\pmb r}-{\pmb r}^j|} \ ,
	\end{eqnarray}
	where ${\pmb n}=(n_1,n_2,n_3)$ are the integer lattice vectors and  ${\pmb k}=(\frac{2\pi}{L}n_1,\frac{2\pi}{L}n_2,\frac{2\pi}{L}n_3)$ are the reciprocal lattice vectors.   The damping parameter $\kappa_e$ is chosen so that we can replace the sum over  ${\pmb n}$  by a simple periodic boundary condition for the short range part of the electrostatic potential in the real space.   This
	is possible as long as $\kappa_e>5/L$.  A special care must be taken in evaluating the ${\pmb k}={\pmb 0}$ term~\cite{dos2016simulations}.  Expanding around $|{\pmb k}|=0$ this term can be written as:
	\begin{eqnarray}\label{div}
	\lim_{{\pmb k} \rightarrow 0}\sum_{j=1}^{N}\text{q}^j\frac{1}{|{\pmb k}|^2}-\sum_{j=1}^{N}\text{q}^j\frac{1}{4\kappa_e^2}+\nonumber \\
	\lim_{{\pmb k} \rightarrow 0}\sum_{j=1}^{N}\text{q}^j\frac{i{\pmb k}\cdot({\pmb r}-{\pmb r}^j)}{|{\pmb k}|^2} -
	\lim_{{\pmb k} \rightarrow 0}\sum_{j=1}^{N}\text{q}^j\dfrac{[{\pmb k}\cdot({\pmb r}-{\pmb r}^j)]^2}{2|{\pmb k}|^2} \ .
	\end{eqnarray}
	The first term is divergent, however it is multiplied by $\sum_i \text{q}^i $ which for a charge neutral system is zero.  Similarly it is possible to show that the third term is also zero by symmetry~\cite{dos2016simulations}.  The only non-trivial term is the last one which evaluates to a finite value, resulting in electrostatic potential at position ${\pmb r}$ inside the simulation cell given by
	\begin{eqnarray}\label{e6}
	\phi({\pmb r})&=&\sum_{{\pmb k \neq 0}}^{\infty}\sum_{j=1}^{N}\frac{4\pi \text{q}^j}{\epsilon_w V |{\pmb k}|^2}\exp{[-\frac{|{\pmb k}|^2}{4\kappa_e^2}+i{\pmb k}\cdot({\pmb r}-{\pmb r}^j)]}  \nonumber \\
	&&+\sum_{j=1}^{N}\text{q}^j\sum_{{\pmb n}}\frac{\text{erfc}(\kappa_e|{\pmb r}-{\pmb r}^j-L{\pmb n}|)}{\epsilon_w |{\pmb r}-{\pmb r}^j|} \nonumber \\
	&& -\sum_{j=1}^{N}\dfrac{2\pi \text{q}^j }{3\epsilon_w V  } ({\bf r}-{\bf r}^j)^2 \hspace{1cm}\ . \hspace{1cm}
	\end{eqnarray}
 We can recognize the last term of this expression as the sum over electrostatic potentials produced by infinite uniformly charged spheres -- each with charge density $q_i/V$  --  centered on positions of ions.  This provides us with an interesting interpretation of Ewald summation.  Effectively it replaces each ion, and its respective replicas,  by infinite uniformly charged spheres centered on positions of physical ions.  The discreteness effects are then encoded in the first two terms of Eq. (\ref{e6}) which correspond to ions inside a neutralizing background.  Note that this interpretation applies also to charge non-neutral systems~\cite{dos2016simulations}.

  For charge neutral system we can rewrite  expression (\ref{e6}) as
	\begin{eqnarray}\label{e6bb}
	\phi({\pmb r})&=&\sum_{{\pmb k \neq 0}}^{\infty}\sum_{j=1}^{N}\frac{4\pi \text{q}^j}{\epsilon_w V |{\pmb k}|^2}\exp{[-\frac{|{\pmb k}|^2}{4\kappa_e^2}+i{\pmb k}\cdot({\pmb r}-{\pmb r}^j)]}  \nonumber \\
	&&+\dfrac{4\pi }{3\epsilon_w V  }  {\bf r} \cdot {\bf M}-\sum_{j=1}^{N}\dfrac{2\pi \text{q}^j \pmb{r}^j\cdot\pmb{r}^j}{3\epsilon_w V  } \hspace{1cm} \nonumber \\
	&&+\sum_{j=1}^{N}\text{q}^j\sum_{{\pmb n}}\frac{\text{erfc}(\kappa_e|{\pmb r}-{\pmb r}^j-L{\pmb n}|)}{\epsilon_w |{\pmb r}-{\pmb r}^j|} \ . \hspace{1cm}
	\end{eqnarray}
	where $ {\pmb  M}=\sum_{j=1}^N  \text{q}^j {\pmb r}^j $ is the electric moment of the simulation cell with $N$ ions and the sum over the  short range interaction is performed using simple periodic boundary condition.
	We recognize the ${\bf r} \cdot {\bf M}$ term as the shape dependent surface potential produced by a macroscopic ferroelectric~\cite{ballenegger2009simulations,Ballenegger}.  The surface term is  particularly important for systems with broken symmetry, such as slab geometry and  ion channels~\cite{smith1981electrostatic,yeh1999ewald,dos2016simulations,yi2017note,telles2021simulations}.  For spherically symmetric bulk systems this term, however, leads to an unrealistic net dipole moment of a macroscopic system, which is clearly absent in the disordered state of an electrolyte solution.  We can remove this term by using tin-foil boundary condition in which our macroscopic system is enclosed by a perfect conductor~\cite{Ballenegger,Leeuw}.  Indeed, as we will show in the following sections,  expression without the surface term results in a better agreement with the Mean Spherical Approximation, which is exact at infinite dilution.   
	The electrostatic energy of a charge neutral system with $N$ ions is then:
	\begin{eqnarray}\label{e6b}
	U_N&=&\frac{1}{2}\sum_{j=1}^N \text{q}^j \left[\phi({\bf r}^j)-\lim_{{\bf r}\rightarrow {\bf r}^j}\frac{q^j} {|{\bf r}-{\bf r}^j|}\right] .\hspace{1cm}
	\end{eqnarray}
	Using Eq. (\ref{e6bb}), this can be written as:
	\begin{eqnarray}\label{Ut}
	&&U_N=\sum_{{\pmb k}\neq{\pmb 0}}^{\infty}\frac{2\pi}{\epsilon_w V |{\pmb k}|^2}\exp{[-\frac{|{\pmb k}|^2}{4\kappa_e^2}]}[A({\pmb k})^2+B({\pmb k})^2]+ \nonumber \\	
	&&\dfrac{1}{2}\sum_{i \ne j}^N\text{q}^i\text{q}^j\frac{\text{erfc}(\kappa_e|{\pmb r}^i-{\pmb r}^j|)}{\epsilon_w |{\pmb r}^i-{\pmb r}^j|} +\dfrac{2\pi }{3\epsilon_w V  }  |{\bf M}|^2- \frac{\kappa_e}{\sqrt{\pi}} \sum_{i=1}^N (\text{q}^i)^2, \ \hspace{1cm}
	\end{eqnarray}
	where
	\begin{eqnarray}
	A({\pmb k})=\sum_{i=1}^N \text{q}^i\text{cos}({\pmb k}\cdot{\pmb r}^i) \ , \nonumber \\
	B({\pmb k})=-\sum_{i=1}^N \text{q}^i\text{sin}({\pmb k}\cdot{\pmb r}^i) \ . \nonumber \\
	\end{eqnarray}
	This is the electrostatic energy for the vacuum boundary condition, in which the surface ${\bf M}$ appears explicitly.  On the other hand, the tin-foil boundary condition entail removal of the $ |{\bf M}|^2$ term from Eq.~(\ref{Ut})~\cite{Ballenegger,Leeuw}.  
	We now compare  $\mu^{\text{ex}}_{\text{t}}$ calculated using the GCMC with vacuum and the tin-foil boundary conditions for symmetric 1:1 electrolyte, with the theoretical result obtained using  the Mean Spherical Approximation (MSA)  with Carnahan-Starling expression for the
	excluded volume interaction.   The MSA+CS expression, $\mu^{\text{ex}}_{\text{t}}=\mu_{\text{MSA}}+ \mu_{\text{CS}}$, is exact for dilute electrolyte~\cite{ho1988mean,ho2003interfacial,levin1996criticality,waisman1972mean,waisman1972mean2,blum1975mean,carnahan1969equation,carnahan1970thermodynamic,adams1974chemical,maciel2018chemical} with:
	\begin{eqnarray}\label{MSA}
	&&\mu_{\text{MSA}}= \frac{\lambda_B\left( \sqrt{1+2 \kappa d}-\kappa d -1\right)} {d^2\kappa}, \\
	&&\mu_{\text{CS}}= \frac{8\eta-9 \eta^2+3\eta^3}{\left(1-\eta\right)^3},
	\end{eqnarray}
	where $\eta=\frac{\pi d^3}{3} c_t$,  $d$ is the ionic diameter,  $c_t=c_++c_-$ is the total concentration of ions, and  $\kappa= \sqrt{8 \pi \lambda_B c_t}$ is the inverse Debye length.
	In simulations we use a cubic cell of length 100 \AA. 
	\begin{figure}[H]
		\centering
		\includegraphics[width=0.7\linewidth]{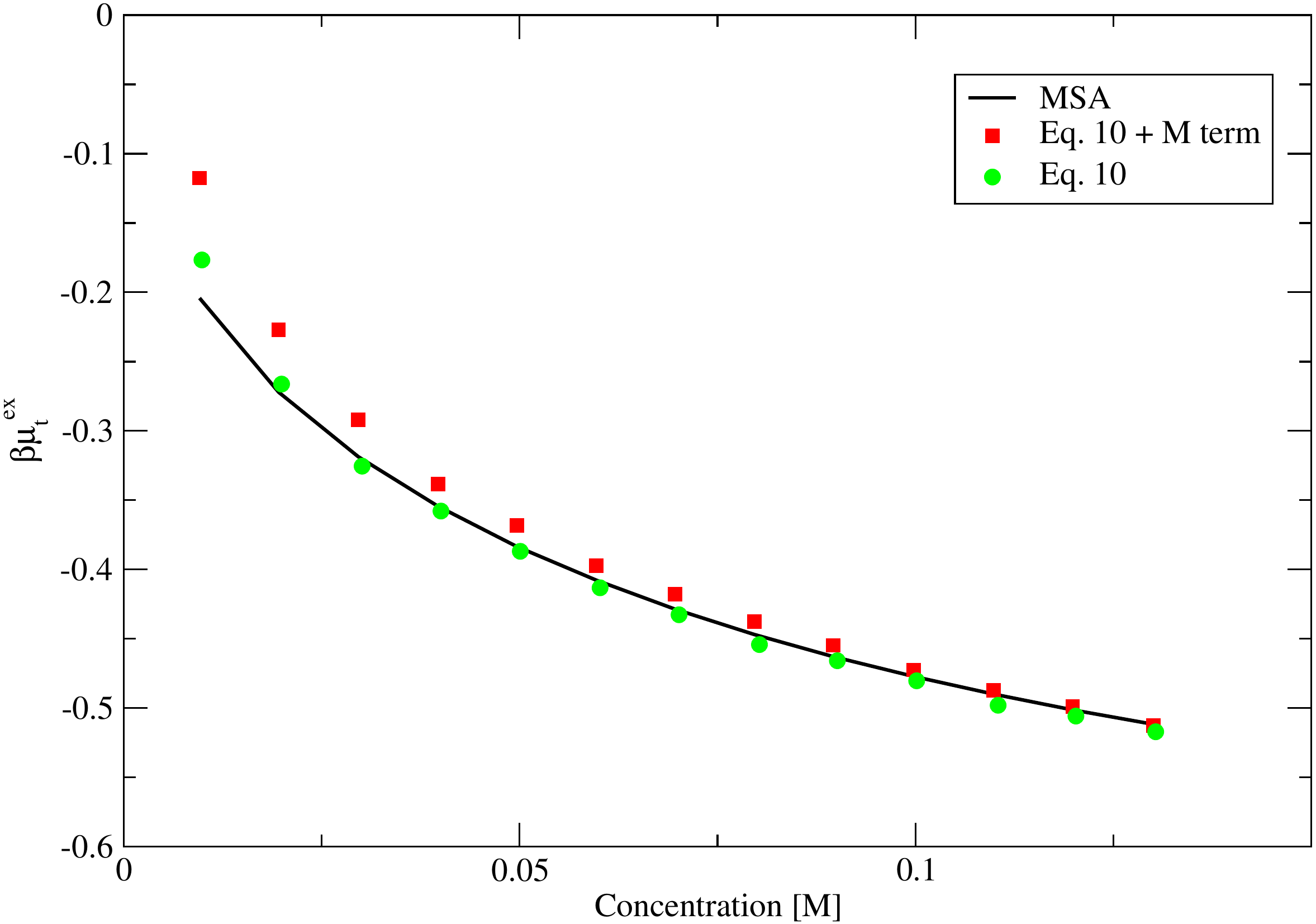}
		\caption{Total excess chemical potential of symmetric 1:1 electrolyte calculated using GCMC simulations with electrostatic energy given by Eq.~\ref{Ut} with M term  (vacuum boundary condition) and without M term (tin-foil boundary condition), compared with the theoretical MSA+CS result. }
		\label{gcmc}
	\end{figure}
	As expected, the  Fig.~\ref{gcmc} shows that Eq.~\ref{Ut} with tin-foil boundary condition results in a better agreement with the theoretical curve at low concentrations of electrolyte. For larger simulation cells the difference between vacuum and tin-foil boundary condition becomes less important.

	\subsection{Method I}
	
	As discussed previously, we have two options for implementing the Widom insertion in a system with Ewald summation. In the first approach we simply insert a new ion of charge $Q$ at position ${\pmb r}^i$.  The change in the electrostatic energy due to the interaction of this ion with all the other ions inside the system and with their replicas is then:
	\begin{eqnarray}\label{deltaphi}
	\Delta U =  Q \phi(\pmb{r}^i)  
	\end{eqnarray}
	where $\phi(\pmb{r}^i)$ is the electrostatic potential at position of insertion given by Eq. (\ref{e6bb}) without the ${\bf M}$ term for tin-foil boundary condition. 
	
	\subsection{Method II}
	An alternative approach is to treat the inserted ion on the same footing as all the other ions inside the simulation cell -- replicating it, along with all the other ions.  In this case the inserted ion will also interact with its own replicas, leading to a diverging electrostatic energy.  The divergence appears in the first term of the expression (\ref{div}), which is no longer zero, since there is a net charge inside the simulation cell.  To overcome this difficulty we  introduce, together with the
	test ion of charge $Q$,  a uniform neutralizing background of opposite charge density $\rho_b(\pmb{r})=-Q/V$, which will also be replicated together with the ions.  Any periodic 
	density function over a cubic lattice can be written as 
	\begin{eqnarray}
	\rho(\pmb{r})=\frac{1}{V}\sum_{\pmb{k}} \tilde \rho  (\pmb{k}) \mathrm{e}^{i\pmb{k}\cdot \pmb{r}}\ ,
	\end{eqnarray} 
	with a similar expression for the electrostatic potential.
	The Fourier transforms of the electrostatic potential and of the charge density are:
	\begin{eqnarray}\label{ep_1}
	\tilde  \phi (\pmb{k}) = \int_V \phi (\pmb{r}) \mathrm{e}^{-i \pmb{k}.\pmb{r} }d^3r, \nonumber \\
	\tilde \rho  (\pmb{k}) =\int_V  \rho  (\pmb{r} ) \mathrm{e}^{-i \pmb{k}.\pmb{r} }d^3r,\\
	\nonumber
	\end{eqnarray}
	where $V$ is the volume of the simulation cell.  In particular, for a uniform background charge density we obtain $\tilde \rho_b  (\pmb{k})=-Q \delta_{\pmb{k}, {\pmb 0}}$, where $\delta$ is Kronecker delta. 
	The electrostatic potential produced by the background satisfies the Poisson equation
	\begin{eqnarray}\label{ep}
	\nabla^2 \phi (\pmb{r}) = -\frac{4 \pi \rho  (\pmb{r})}{\epsilon_w } \ .
	\end{eqnarray}
	Substituting the Fourier representation of electrostatic potential and of charge density into Eq. (\ref{ep}) we obtain
	\begin{eqnarray}\label{ep_2}
	\tilde \phi(k) = \frac{4 \pi}{\epsilon_w}\frac{\tilde \rho (\pmb{k})}{ k^2 } .
	\end{eqnarray}
	Finally, using the expression for the Fourier transform of the uniform background charge, we obtain the 
	contribution that it produces to the total electrostatic potential:
	\begin{eqnarray}\label{back}
	\phi_b(\pmb{r}) =-\frac{4 \pi Q}{V \epsilon_w} \sum_{\pmb{k}} \mathrm{e}^{i \pmb{k}.(\pmb{r} -\pmb{r}^i)} \frac{\delta_{\pmb{k},0}}{ k^2},
	\end{eqnarray}
	where we have centered the background on the position of the inserted ion.  
	Adding this background potential to the potential produced by all $N+1$ replicated ions we see that the 
	divergence in the  $\pmb{k}={\pmb 0}$  term in expression (\ref{div}) cancels exactly.  There is, however, now an additional term  coming from the $\pmb{k}\rightarrow {\pmb 0}$ limit of Eq. (\ref{back}).  This term is proportional to $(\pmb{r} -\pmb{r}^i)^2$,  and will cancel the same term in Eq. (\ref{e6}) for $N+1$ particle system,  resulting in the total electrostatic potential of a system with a neutralizing background 
	\begin{eqnarray}\label{phib}
	\varphi({\pmb r})&=&\sum_{{\pmb k \neq 0}}^{\infty}\sum_{j=1}^{N+1}\frac{4\pi \text{q}^j}{\epsilon_w V |{\pmb k}|^2}\exp{[-\frac{|{\pmb k}|^2}{4\kappa_e^2}+i{\pmb k}\cdot({\pmb r}-{\pmb r}^j)]}  \nonumber \\
	&&-\sum_{j=1}^{N}\dfrac{2\pi \text{q}^j }{3\epsilon_w V  } ({\bf r}-{\bf r}^j)^2 \hspace{1cm} \nonumber \\
	&&-\frac{Q}{\epsilon_w V\kappa_e^2}+\sum_{j=1}^{N+1}\sum_{{\pmb n}}\text{q}^j\frac{\text{erfc}(\kappa_e|{\pmb r}-{\pmb r}^j -L{\pmb n} |)}{\epsilon_w |{\pmb r}-{\pmb r}^j|}, \ \hspace{1cm}
	\end{eqnarray}
	where we have defined the $j=N+1$ as our test ion with the charge $q_{N+1}=Q$.  Note that the second sum in Eq. (\ref{phib}) runs only over the original ions present in the system.  
	
	Suppose we insert a test ions at position ${\pmb r}^i$, together with the associated neutralizing background, into an initially empty simulation cell, $N=0$.  The electrostatic energy of this system will be:
	\begin{eqnarray}\label{U00}
	U_0=\frac{Q}{2} \lim_{{\pmb r} \rightarrow {\pmb r}^i} \left( \varphi({\pmb r})-\frac{Q}{|{\pmb r}-{\pmb r}^i|}\right) \ .
	\end{eqnarray}
	Performing the limit we obtain 
	\begin{eqnarray}\label{U0}
	\beta U_0=-1.418648739\frac{\alpha^2 \lambda_B}{L}\ ,
	\end{eqnarray}
	where $\alpha$ is the valence of ion of charge $Q=\alpha q$, where $q$  is the proton charge,  and $\lambda_B= {q^2}/{ \varepsilon_w \  k_{\rm B} T} $ is the Bjerrum length.   Note that $U_0$ does not depend on the damping parameter $\kappa_e$.  Eq. (\ref{U0}) is the Madelung energy of a simple cubic lattice of ions of charge $Q$ in a neutralizing background.   It is important to keep in mind that Ewald sums are conditionally convergent and that the background is assumed to be spherically symmetric with respect to the position of the inserted ion.  The energy $U_0$ contains the electrostatic self energy of the background, the interaction energy of ion with the background, and the interaction energy of ion with all of its images. 
	
	The change in electrostatic energy of a charge neutral system with $N$ ions due to the introduction of a replicated test ion at position ${\pmb r}^i$ and a spherical neutralizing background centered on this ion is:
	\begin{eqnarray}\label{du}
	\Delta U= Q \phi({\pmb r}^i)+\frac{2\pi Q }{3\epsilon_w V  }\sum_{j=1}^{N}\text{q}^j  ({\pmb r}^i-{\pmb r}^j)^2 + U_0\ .
	\end{eqnarray}
	The first term in this expression is due to the interaction of ion $Q$, inserted at positions ${\pmb r}^i$, with the $N$ ions of the original charge neutral system and with their replicas.  The electrostatic potential $\phi({\pmb r}^i)$ is given by Eq. (\ref{e6}).   The second  term is the interaction energy of the original $N$ ions with the spherical neutralizing background centered on the inserted ion.  The resulting quadratic potential results in a linear force produced by the background on each ion.   The last term is the interaction energy of the ion $Q$ with its neutralizing background, with its own replicas, as well as the self energy of the neutralizing background.  
	
	The expression can be simplified yielding:
	\begin{eqnarray}\label{du1}
	&&\Delta U=Q\sum_{{\pmb k \neq 0}}^{\infty}\sum_{j=1}^{N}\frac{4\pi \text{q}^j}{\epsilon_w V |{\pmb k}|^2}\exp{[-\frac{|{\pmb k}|^2}{4\kappa_e^2}+i{\pmb k}\cdot({\pmb r}^i-{\pmb r}^j)]}  \nonumber \\
	&&+Q \sum_{j=1}^{N}\text{q}^j\frac{\text{erfc}(\kappa_e|{\pmb r}^i-{\pmb r}^j|)}{\epsilon_w |{\pmb r}^i-{\pmb r}^j|} +U_0.\ \hspace{1cm}
	\end{eqnarray}
	It is interesting to note that this expression does not depend on ${\bf M}$ for either vacuum or tin-foil boundary condition.  This is the case only if the neutralizing background is centered on the inserted ion.
	
	\section{Result} \label{s3}
	
	We now compare the predictions of the two Widom insertion methods with the $\mu_t$  obtained using GCMC simulations.  
	As was discussed in the introduction,  GCMC does not give us individual chemical potentials of ions, but only the value of  $\mu_t$,  which we will use as a benchmark to judge the accuracy of the two Widom insertion methods.
	
\begin{figure}[H]
		\centering
		\includegraphics[width=0.7\linewidth]{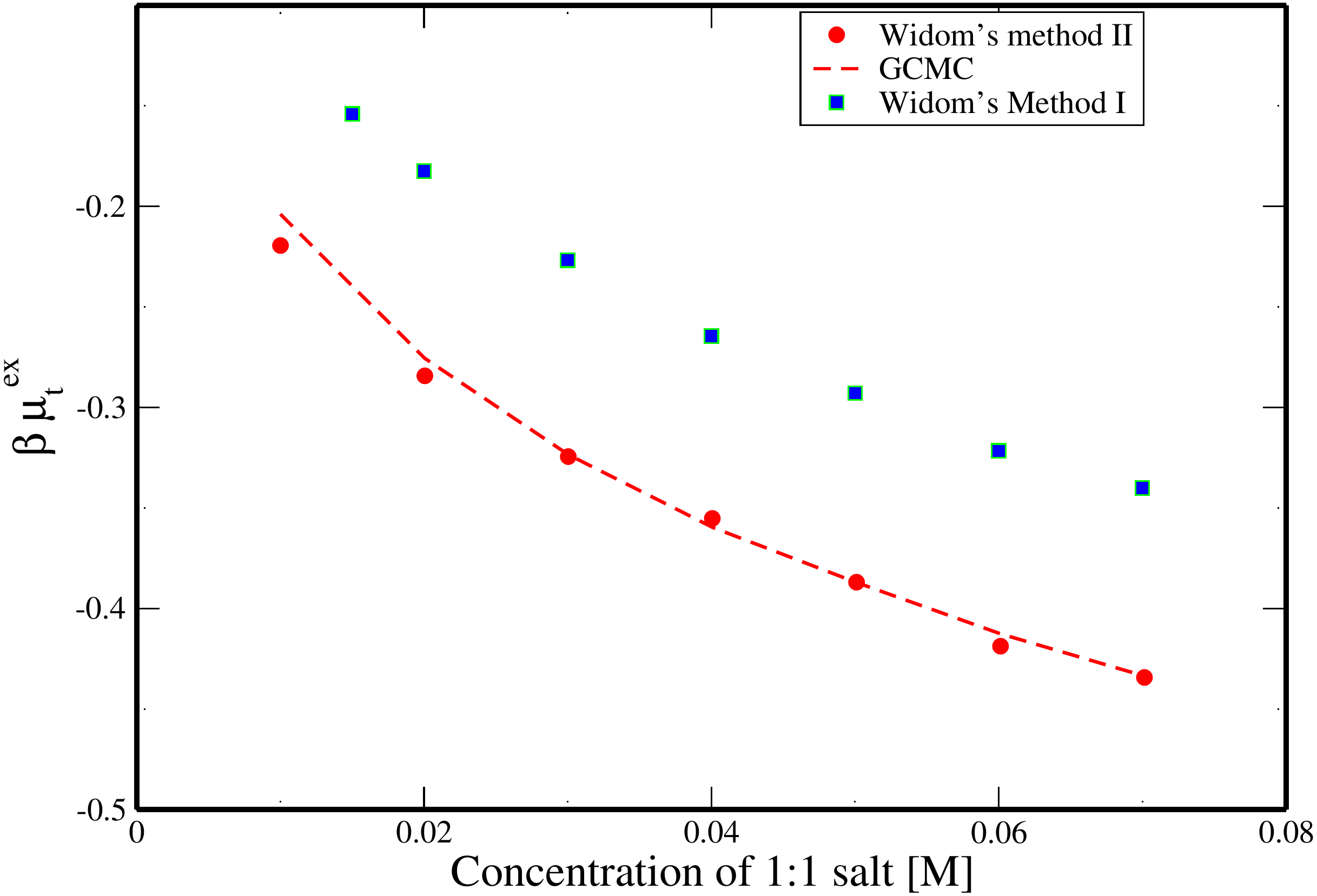}
		\caption{Comparison of the total  excess chemical potential  $\mu_t^{ex}$ obtained using  two Widom insertion methods and the GCMC simulations.   The results of Method I are shown with squares and Method II with circles.  The radii of positive and negative ions are 2~\AA. }
		\label{fg1}
	\end{figure}

	We start with symmetric $1$:$1$ electrolyte.  In Fig~\ref{fg1} we present the $\mu_t^{ex}= \mu_+^{ex} + \mu_-^{ex}$=2$\mu_+^{ex}$=2$\mu_-^{ex}$, obtained using 
	the two Widom insertion methods, compared with the results obtained using the GCMC.  
	
\begin{figure}[H]
		\centering
		\includegraphics[width=0.7\linewidth]{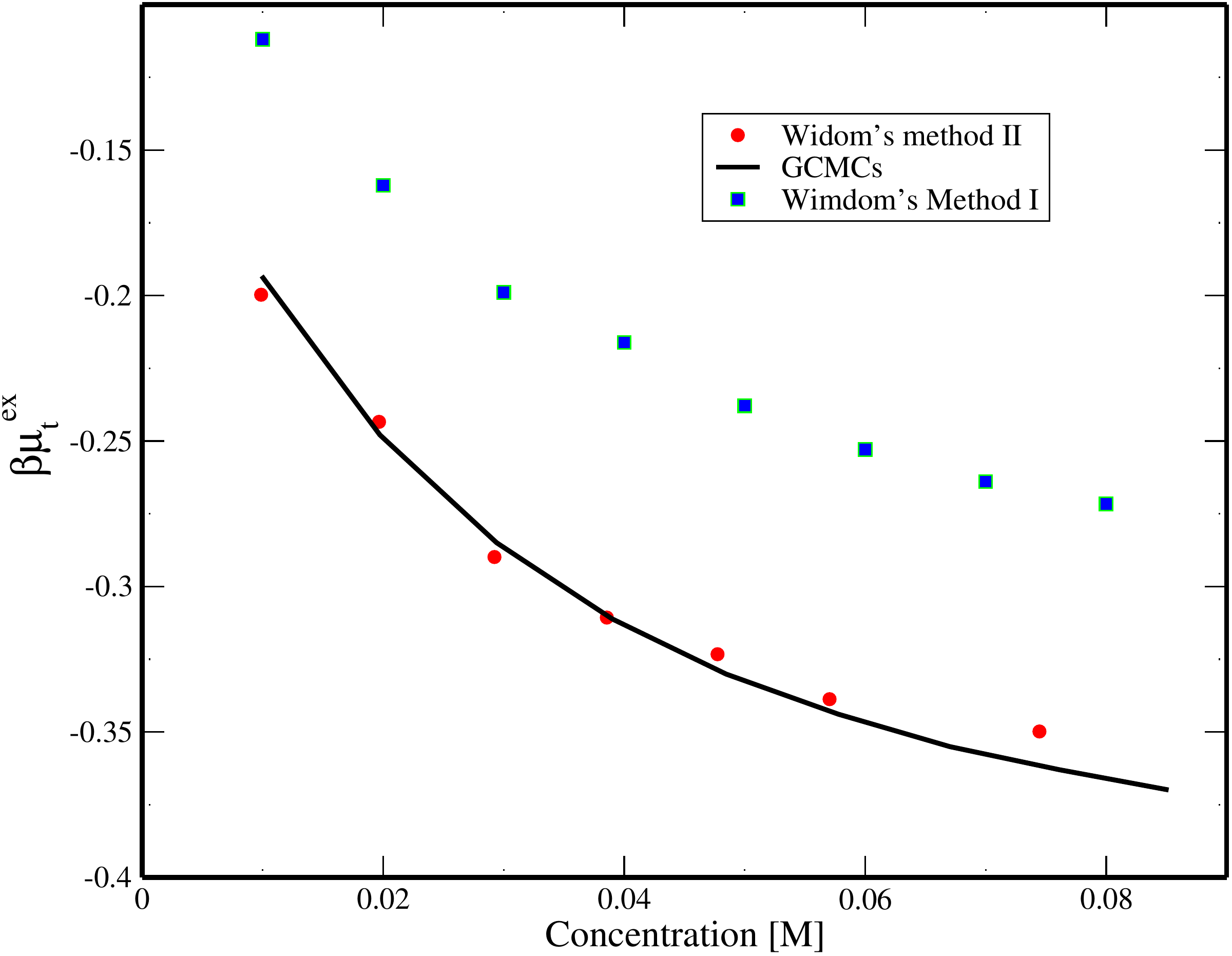}
		\caption{Comparison of $\mu_t^{ex}$ obtained using  Method II with the GCMC  for  1:1 electrolyte  with cations of  radius 2~\AA\  and anions of  3~\AA.}
		\label{fg2}
	\end{figure}

\begin{figure}[H]
		\centering
		\includegraphics[width=0.7\linewidth]{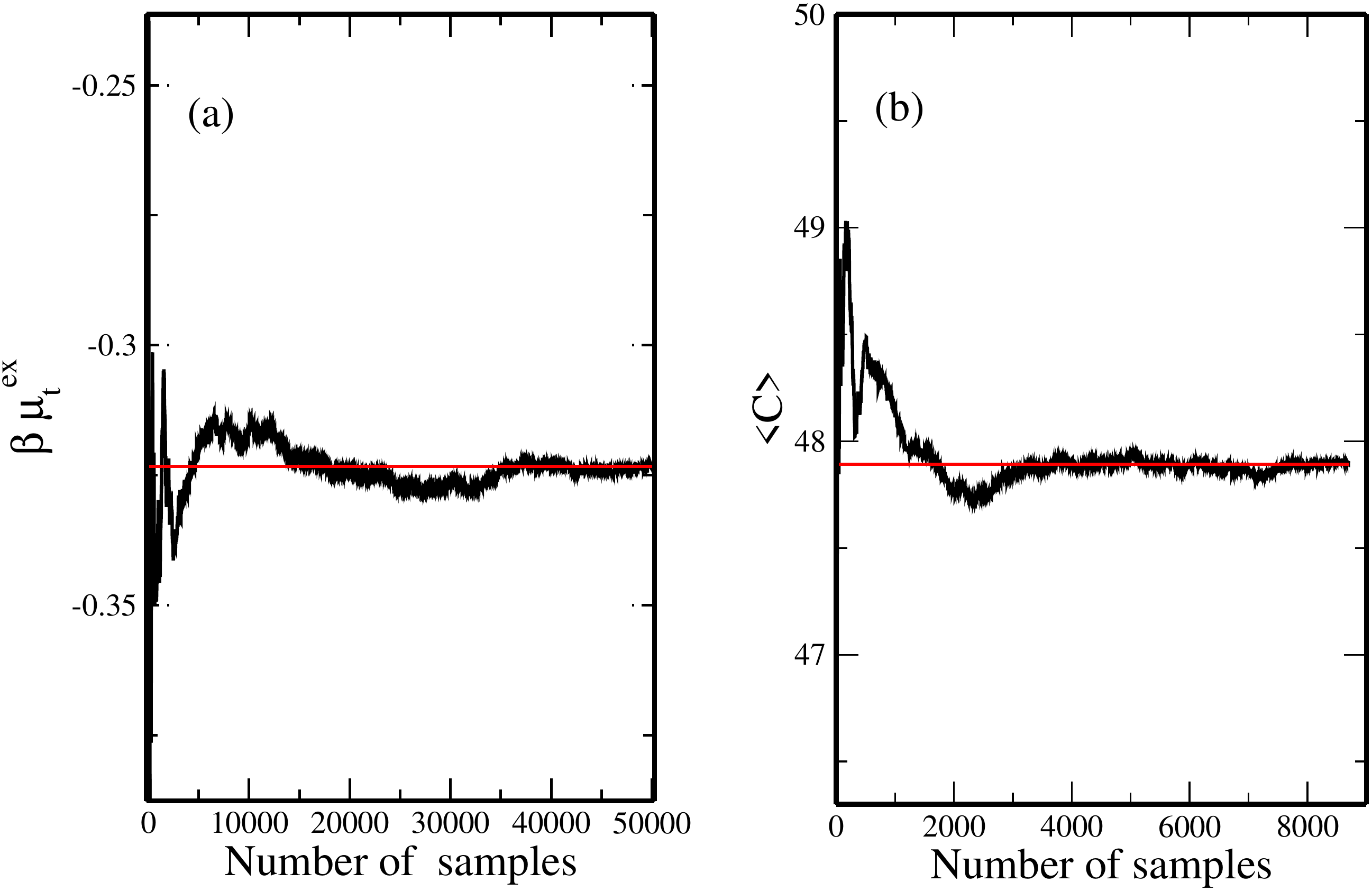}
		\caption{(a) Convergence of the chemical potential obtained using Method II,  for 1:1 electrolyte with cations of  radius 2~\AA\ and anions of  3~\AA\ at concentration of $48$ mM,  as a function of the number of samples used. The $\beta \mu_t^{ex}$ converges to $-0.32$.   In panel (b)  convergence of electrolyte concentration in mM, as a function of samples using GCMC simulation with fugacity fixed at $5.9820 \times 10^{-10}$~\AA$^{-9}$.  With this value we obtain  $\beta \mu_t^{ex}=-0.32$ and the concentration $47.4$mM.  We see that 
			convergence is much faster for GCMC than for Widom insertion, both in terms of the CPU time, since one can use a smaller simulation cell, and also in terms of the number of samples needed to calculate the averages.}
		\label{fs}
	\end{figure}

\begin{figure}[H]
		\centering
		\includegraphics[width=0.7\linewidth]{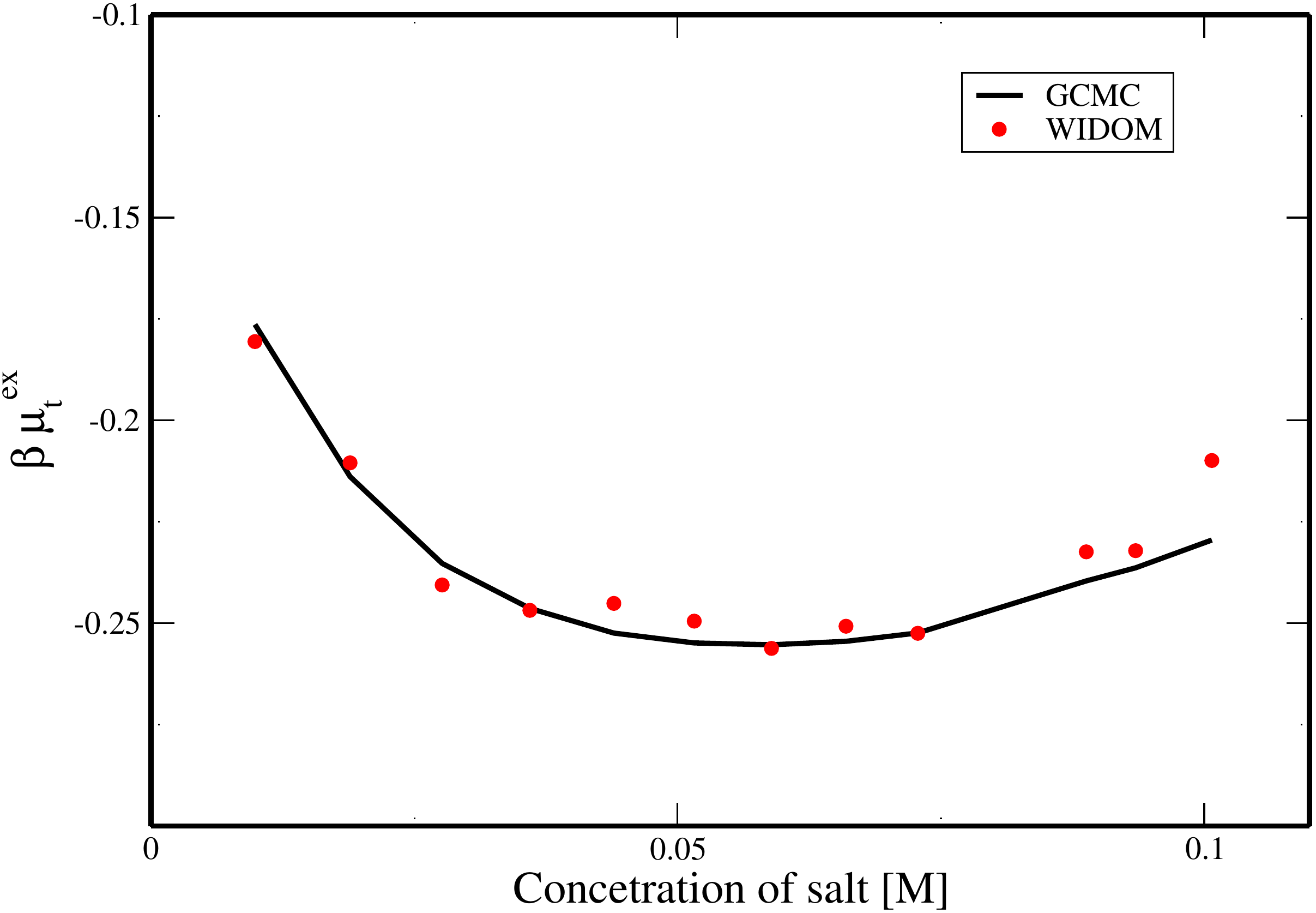}
		\caption{Comparison of $\mu_t^{ex}$ obtained using Method II with the GCMC sumulation results for  1:1 electrolyte  with cation of  radius 2~\AA\ and anions of  4~\AA.  For large size asymmetry between cations and anions the chemical potential is no longer a monotonic function of electrolyte concentration. }
		\label{f2}
	\end{figure}

\begin{figure}[H]
		\centering
		\includegraphics[width=0.7\linewidth]{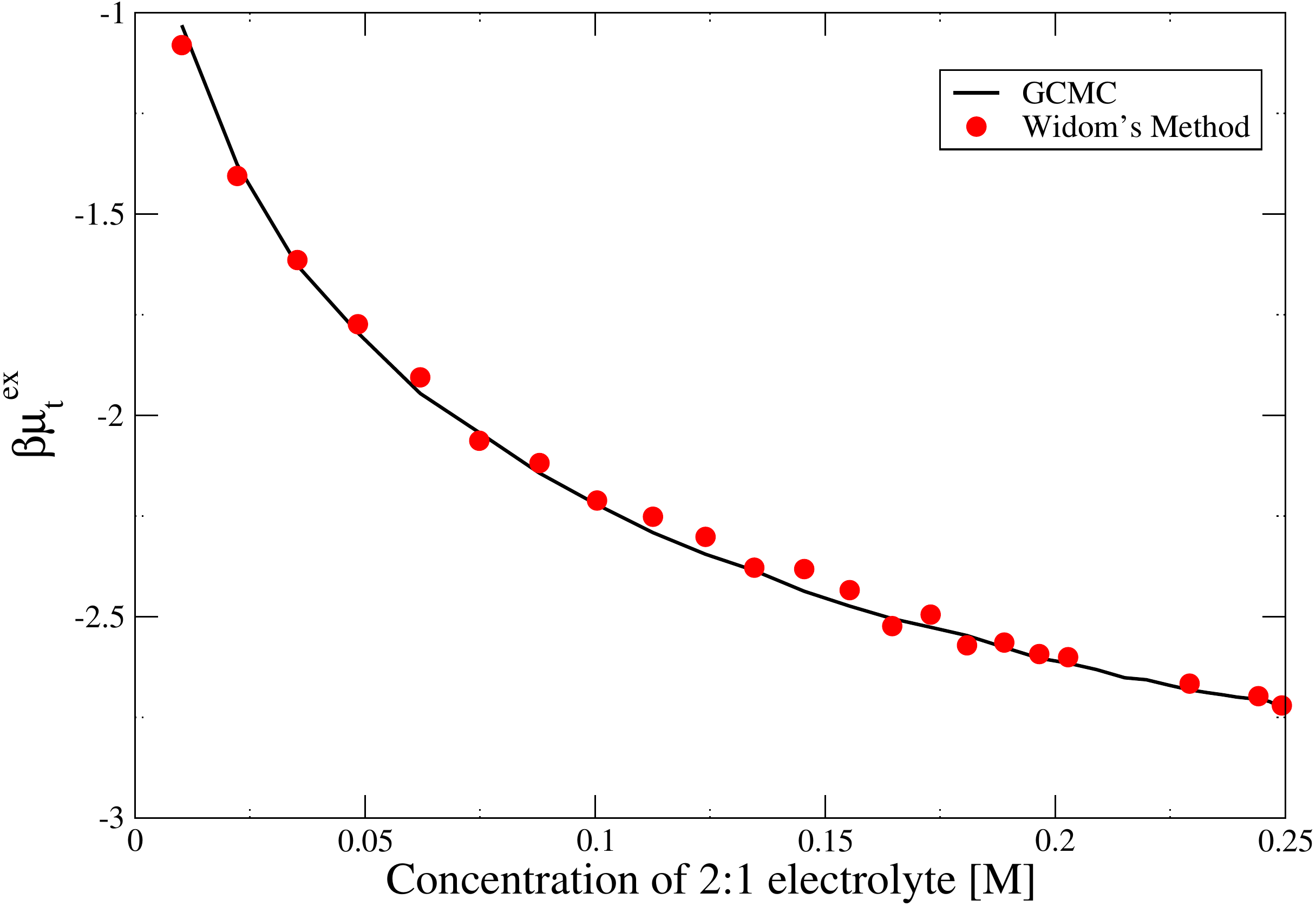}
		\caption{Comparison of the excess chemical potential $\mu_t^{ex}$ obtained using  GCMC simulations with Widom Method II,  for  size symmetric 2:1 electrolyte with ions of radius 2~\AA.}
		\label{f3}
	\end{figure}

	We see that Method I results in a very significant  deviation from  the benchmark GCMC simulation results, while Method II  is in good agreement.  Nevertheless, we see that even for  a fairly large simulation cell of $L=200$~\AA, we have a significant scatter in the data points even after using $50000$ samples to perform averages.  On the other hand, we obtain a smooth curve using GCMC already with $L=200$~\AA~ and  only $10000$ samples.   In fact, with GCMC we obtain the same results even with a much smaller simulation cell of   $L=100$~\AA. We next repeat the calculations for asymmetric 1:1 electrolyte, with cations of radius $2$~\AA~ and anions of radius $3$~\AA.   In Fig. \ref{fg2}(a) we compare the values of  $\mu_t^{ex}$  obtained using the two Widom methods with the ones obtained using the GCMC.   Once again we see that Method II is in much better agreement with the GCMC result.    In Fig. \ref{fs}(a) we show the slow convergence of the Widom insertion method as a function of the number of samples and in Fig \ref{fs} (b) we show the convergence of GCMC.  In the case of GCMC we have fixed the fugacity and calculate the average number of particles inside the simulation cell from which we obtain the average concentration.  The convergence is much faster for GCMC than for Widom insertion.  As we increase the size asymmetry between cations and anion even further, the excess chemical potentials become  non-monotonic functions of concentration, see Fig. \ref{f2}.  The reasonably good agreement between Method II and GCMC still persists, but the Widom data becomes more noisy for the same number of samples.  Finally,  in Fig. \ref{f3} we compare the Method II with GCMC for  size symmetric 2:1 electrolyte, with  ions of  radius $2$~\AA.  In this case $\mu_t^{ex}=\mu_+^{ex}+2 \mu_-^{ex}$.  Again we see 
	a good agreement between GCMC and Method II.
	
	To more clearly see the degree of agreement between Widom insertion method and GCMC,  in Table 1 we present the individual  chemical potentials of cations and anions of 2:1 electrolyte calculated using Method II.  We also compare the resulting values of $\mu_t$ with the ones obtained using GCMC.  We see that even with $50000$ samples the agreement is only to two significant figures.  
	
	\begin{table*} 
		\caption{Individual and total chemical potentials obtained using Method II compared with GCMCs results for 2:1 electrolyte.}\label{tab1}
		\footnotesize
		\setlength{\tabcolsep}{30pt}
		\begin{tabular}{ccccc}
			\hline 
			c [mM] &  $\beta\mu_{\text{++}}$  &   $\beta\mu_{\text{-}}$   & $\beta\mu_{\text{++}} + 2\beta\mu_{\text{-}}$   &    $\beta\mu_{\text{t}}$    \\ 
			\hline
			$10$ & $-0.700$ &  $-0.189$ &$-1.080$   & $-1.030$ \\ 
			
			$60$ & $-1.220$ & $-0.342$  &  $-1.905$  & $-1.945$ \\ 
			
			$100$ & $-1.439$ & $-0.386$ &  $-2.211$&  $-2.222$  \\ 
			
			$145$ &  $-1.567$ & $-0.407$ & $-2.381$ &$-2.437$  \\ 
			
			$172$ & $-1.651$ & $-0.421$ &$ -2.495 $     &  $-2.526   $ \\ 
			
			$202$ & $-1.739$ & $-0.430 $ & $ -2.600$ &  $-2.615$ \\ 
			
			$252$ &  $-1.834$ & $-0.433$ & $-2.701$   &  $-2.722$\\ 
			\hline
		\end{tabular} 
	\end{table*}

	\section{Conclusion} \label{s4}
	We have explored the use of Widom insertion method for calculating the chemical potential of individual ions in computer simulations with Ewald summation.  Two approaches were considered.  In the first approach an individual ion is  inserted into a periodically replicated overall charge neutral system representing an electrolyte solution.  In the second approach, an inserted ion is also periodically replicated, resulting  in a macroscopic  violation of the overall charge neutrality.   To overcome this problem, a neutralizing background must be introduced simultaneously  with the ion.   This results in a linear force that background  exerts on all the ions.  Comparing the results of the two methods, we find that the second approach is in much better agreement with the benchmark GCMC simulations for the total chemical potential of the ions $\mu_t$.  This is consistent with the results obtained using the minimum image simulations, which were also found to require a neutralizing background to improve convergence~\cite{sloth1990monte,malasics2010efficient}, as well as with the simulations of ionic solvation~\cite{hummer1996free}.    We find that to be accurate,  Widom insertion method requires very large simulation cells.  Apparently only for very large cells the contribution of background to the chemical potential becomes negligible.   To produce  reasonably accurate values of the chemical potential of individual ions,  a very large number of samples must also be used.  Therefore, in applications which do not require knowledge of  the individual ionic chemical potentials, but only of $\mu_t$,  the GCMC approach is by far more practical.   
	
	  The significant difference between Widom I and Widom II methods is quite surprising.  Its origin can be traced back to the careful limit of the $k=0$ term of Ewald potential, see Eqs. (\ref{e6})  and (\ref{e6bb}).  The limit results in a term quadratic  in ion positions, as well as  ${\bf M}$  dependent contribution, Eq. (\ref{e6bb}).  These terms are usually neglected appealing to tin-foil boundary condition.   However, tin-foil will only removes the $M$ dependent term, while the quadratic term still remains.   Indeed,  the quadratic term is of fundamental importance when studying  non-neutral systems such as
ions confined between like charged plates, see for example Ref. \cite{dos2016simulations}.  It is precisely the quadratic term  that leads to the deviation between Widom I and II.  When using Widom II, the quadratic term cancels precisely by the interaction with a neutralizing background that is introduced together with the inserted ion, see Eq. (\ref{du}).
	
	\section*{Acknowledgments}
	This work was partially supported by the CNPq, CAPES and National Institute of Science
	and Technology Complex Fluids INCT-FCx. The authors are grateful to the Instituto de
	Física e Matemática, UFPel, for the use of computer resources.
	\section{Conflict of Interest}
	The authors have no conflicts to disclose.
	\section{DATA AVAILABILITY}
	The data that support the findings of this study are available
	from the corresponding author upon reasonable request.
	\section*{References}

\end{document}